\documentclass[aps,twocolumn,pra,amssymb,amsmath, bm, floatfix,superscriptaddress,nofootinbib,float,10pt]{revtex4}
\usepackage{graphicx}
\usepackage{mathptmx}
\begin{document}
\title{Turbulence in Binary Bose-Einstein Condensates Generated by Highly Non-Linear Rayleigh-Taylor and Kelvin-Helmholtz Instabilities}
\author{D. Kobyakov$^{*}$}
\affiliation{Department of Physics, Ume{\aa} University , SE-901 87 Ume{\aa}, Sweden}
\author{A. Bezett}
\affiliation{Institute for Theoretical Physics, Utrecht University, Leuvenlaan 4, 3584 CE Utrecht, The Netherlands}
\author{E. Lundh}
\affiliation{Department of Physics, Ume{\aa} University , SE-901 87 Ume{\aa}, Sweden}
\author{M. Marklund}
\affiliation{Department of Applied Physics, Division of Condensed Matter Theory, Chalmers University of Technology, SE-412 96 G{\"o}teborg, Sweden}
\author{V. Bychkov}
\affiliation{Department of Physics, Ume{\aa} University , SE-901 87 Ume{\aa}, Sweden\\
$^{*}$dmitry.kobyakov@physics.umu.se}

\begin{abstract}
Quantum turbulence (QT) generated by the Rayleigh-Taylor instability in binary immiscible ultracold $^{87}\texttt{Rb}$ atoms at zero temperature is studied theoretically. We show that the quantum vortex tangle is qualitatively different from previously considered superfluids, which reveals deep relations between QT and classical turbulence. The present QT may be generated at arbitrarily small Mach numbers, which is a unique property not found in previously studied superfluids. By numerical solution of the coupled Gross-Pitaevskii equations we find that the Kolmogorov scaling law holds for the incompressible kinetic energy. We demonstrate that the phenomenon may be observed in the laboratory.
\end{abstract}

\maketitle

\section{Introduction}
Superfluids have become a useful experimental and theoretical tool for investigation of the basic concepts and properties of turbulence, from cascades to intermittency. Many similarities between turbulence in classical and quantum fluids have been found, such as the Kolmogorov spectrum observed experimentally in helium \cite{EPL_43_29_1998}, \cite{EPL_97_34006_2012}, and found theoretically in the Gross-Pitaevskii (GP) model \cite{PRL_94_065302_2005}, the dissipation dynamics measured in helium in the quasiclassical regime of quantum turbulence (QT) \cite{PRL_99_265302_2007}, \cite{NatPhys_7_473_2011}, and strong intermittency in QT \cite{EPL_43_29_1998}, \cite{EPL_98_26002_2012}, \cite{PRL_110_014502_2013}. Recent studies of QT revealed that the vortex bundles are crucial for the Kolmogorov spectrum of QT \cite{Phys_Fluids_24_055109_2012}, \cite{PRA_86_053621_2012}. The bundles are a counterpart of the coherent structures in classical turbulence (CT) \cite{Nat344_226_1990}.

However, strong differences exist between QT and CT, first of all, because of the underlying velocity fields. The quantum counterpart of the classical sheet-like spatial structure of vorticity of the flow, which becomes turbulent due to roll-up caused by the Kelvin-Helmholtz instability (KHI), has been missing in QT so far. The topological charge [i.e.,
vorticity of quantum fluid defined before Eq. (8)], in QT is a tangle of one-dimensional (1D) lines where the phase of the superfluid wavefunction is singular (collection of vortex points in 2D QT). The characteristic vortex core size is $\sim0.1\mu\mathrm{m}$ \cite{PRA_58_4816_1998} in ultracold gases, and $\sim0.1\mathrm{nm}$ in superfluid helium \cite{Donnelly1991}. In 3D QT vorticity decays $\propto t^{-1}$, where $t$ is time, in contrast to $\propto t^{-3/2}$ for the quasiclassical regime \cite{PRL_100_245301_2008}. This is a consequence of a different mechanism of dissipation of incompressible kinetic energy (in other words, the transverse contribution to the kinetic energy) in QT: viscosity is absent, and decrease of the vortex line length happens due to its collisions and reconnections. Another signature of QT is velocity statistics at wavenumbers\footnote{Uniform velocity $\mathbf{v}$ of a superfluid described by a wavefunction $\psi=\sqrt{n}e^{i\mathbf{kr}}$, is $\mathbf{v}=(\hbar/2mi|\psi|^2)(\psi^*\nabla\psi-\psi\nabla\psi^*)=(\hbar/m)\mathbf{k}$, i.e. is proportional to the wavenumber $\mathbf{k}$, where $n$ is number density, and $m$ is atomic mass. } larger than the inverse mean intervortex distance \cite{PRL_101_154501_2008}, \cite{PRL_104_075301_2010}, \cite{PRB_83_132503_2011}. Incompressibility is usually a very good approximation in helium because mean intervortex separation is much larger than the core size \cite{PRB_38_2398_1988}, \cite{PhysRep_522_191_2013}, \cite{PhysRep_524_85_2013}. The compressible effects become important for dynamics on scales comparable to the core size, especially in Bose-Einstein condensates (BEC), with several non-trivial consequences such as sound generation due to reconnections \cite{PRL_86_1410_2001}, thermal dissipation \cite{PRL_97_145301_2006}, \cite{LaserPhysLett8_691}, direct \cite{PRA_81_063630_2010} and inverse \cite{PRL_110_104501_2013} energy cascades in two-dimensional QT, and make QT quite peculiar. It is therefore natural to ask, is it possible to find deeper relations between the superfluid and the classical turbulence using available experimental systems?

Here we address this issue by considering the GP model which describes binary BEC \cite{PethickSmith}. It is worth to note that in
the thermodynamic limit Bose-Einstein condensation can take
place only in 3D [25]. Here we utilize the term “BEC” for
brevity, implying that the entire cloud of atoms is superfluid
and phase-coherent. Contrary to single-component BECs, in
binary BECs close to the SU(2) symmetry considered here
(see Sec. IIA for details), vortex cores of one component
filled by nonrotating atoms of the other component do not
produce significant total density gradients. Thus, in turbulence
the total density does not experience significant perturbations,
the total fluid velocity should be almost incompressible, and
the spectrum of the total quantum kinetic energy is expected
to obey the Kolmogorov scaling. The present work is based on our previous theoretical studies where we investigated the RTI in ultracold $^{87}\mathrm{Rb}$ \cite{PRA_83_043623_2011}, and explored some alternative scenarios \cite{PRA_85_013630_2012}. Here we utilize the same physical model of two
immiscible BECs with two opposite atomic spinswhich makes
possible the exertion of the opposite forces on both components
by an inhomogenous magnetic field, as first suggested by
the authors of [28]. The magnetic field creates excess of the
position-dependent potential energy of each BEC component,
and leads to the RTI.

The atomic magnetic moment together with the twocomponent
character of superfluid, are essential for the
manipulation of the condensate components and open
up the way to stir the superfluid in a controlled way. Generally, the miscibility of superfluids is determined by stability of a uniform mixture with respect to density fluctuations. When the second-order energy perturbation is positive definite, the system is miscible. For such external trapping potentials when quantum pressure of each superfluid in the relevant spatial dimensions can be  neglected (the Thomas-Fermi approximation), the stability condition is equivalent to $g_{11}>0$, and $g_{11}g_{22}>g_{12}^2$ \cite{PethickSmith}, where $g_{ij}$ are the interaction parameters
(see Sec. IIA for details). Thus, immiscible systems satisfy $g_{11}g_{22}<g_{12}^2$. The stirring procedure that we propose, is based on the Rayleigh-Taylor instability (RTI) \cite{PRA_80_063611_2009},\cite{PRA_83_043623_2011} and the KHI \cite{PRB_81_094517_2010}. The former triggers the components mixing, while the latter performs the actual stirring of the superfluids.

We reveal a special role of QT in immiscible\footnote{Using the condensed-matter terminology, immiscible superfluids can be called "ferromagnetic" in the sense of their pseudo-spin vector field $\mathbf{S}$ and its $z$-component: the non-linear spin-dependent contribution to the total energy is $\propto-\gamma S_z^2\equiv-\gamma(n_1-n_2)^2$ \cite{PRA_85_013630_2012}, being negative definite for $\gamma>0$, which implies local maximization of $S_z^2$.} superfluids close to SU(2)-symmetry, which is equivalent to equal atomic masses $m$, equal non-linear interaction parameters $g_{11}=g_{22}$, equal bulk densities of the two BECs, and small interspecies repulsion parameter $\gamma\equiv g_{12}/g-1$. We show in Eq. (\ref{RTI}) and below that the interface tension between the components is proportional to $\sqrt{\gamma}$. Dimension of the topological charge makes binary QT fundamentally different from the previously studied cases for single-component superfluids \cite{PhysRep_524_85_2013}. Although dimension of the vorticity structure of miscible superfluids \cite{JLowTempPhys_162_361_2010}, and immiscible superfluids with strong inter-component repulsion \cite{PRA_85_033642_2012} is the same as of weakly repulsive BECs considered here, the mean of turbulent vorticity of weakly repulsive superfluids is still associated with different structures. Because of immiscibility, sheet-like structures are the most favorable vorticity structures, in contrast to tube-like structures in miscible superfluids. In strongly repulsive superfluids, sheet-like structures may be present, but they are stabilized with respect to the KHI [27], and are not expected to be structures where the mean turbulent vorticity resides. Thus, although still created by quantized phase windings of superfluid wave functions, the topological charge of weakly repulsive BECs is non-quantized (because the interfacial shear motion may be arbitrarily slow), and represents a quantum counterpart of the random vorticity field of CT.

\section{The RTI in binary BEC}
In hydrodynamics of classical fluids, the RTI occurs at the interface between fluids in the gravitational field. In the simplest case the interface is flat and perpendicular to the gravitational force. When a lighter fluid is above a heavier one, waves on the interface are stable, and the spectrum of the capillary-gravitational waves is real for all wavenumbers. However in the opposite situation, fluids tend to swap their locations: the lighter fluid moves up, and heavier fluid falls down lowering the potential energy. In fact, before the swapping the fluid is in the unstable (hydrostatic) equilibrium unless the interface is slightly perturbed. Perturbation wavelength must be sufficiently long to cause the RTI.

In the context of ultracold boson gas the role of gravity can be taken by the magnetic field. The force results from spatially-dependent energy of interaction of atomic magnetic moments of superfluid with the magnetic field, and acts perpendicularly to the interface. The spatial dependence of energy is caused by a spatially uniform $dB_x/dz$, where $B$ is external magnetic field, $x$ is along the interface, and $z$ is perpendicular to the interface.

\subsection{The model}
We focus on 2D dynamics considering two BECs residing in two halves of the harmonic trap. The number of atoms $\sim5\times10^6$, and the scattering lengths $a_{11}=a_{22}=100.4a_B$, where $a_B$ is the Bohr radius, are the same for both BECs. The interspecies scattering length is $a_{12}=101.8a_B$ \cite{PRA_80_063611_2009}, \cite{RMP_85_1191_2013}. Thus, in the ground state the BECs are immiscible, and the interface between them is plane as shows Fig. 1 a.
\begin{figure}
\includegraphics[width=3.55in]{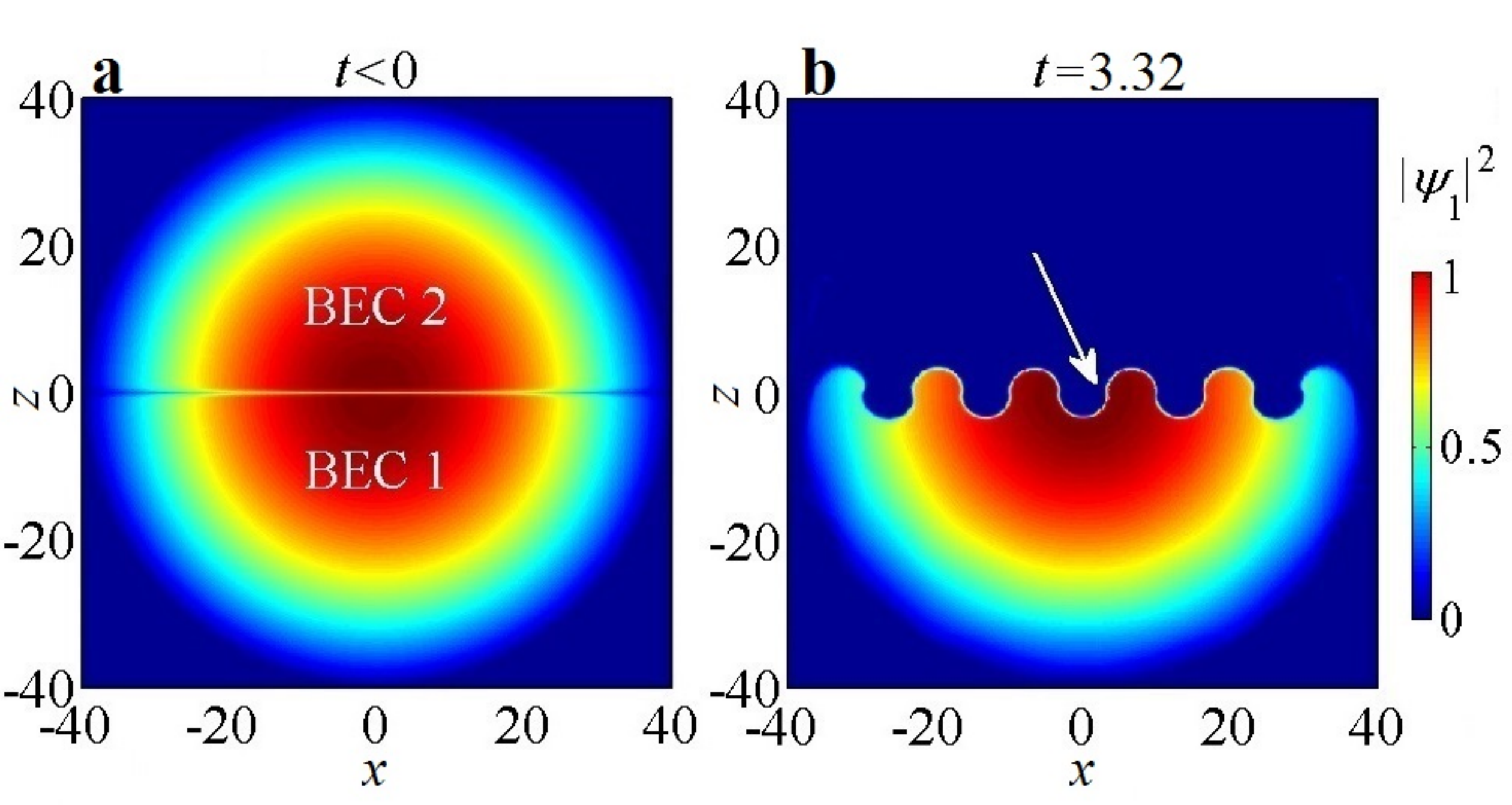}
\caption{(Color online) (a) The ground state of binary BEC with weakly repulsive inter-component interaction in a 2D harmonic trap for $\gamma=0.01$ and $R_0=40$. (b) Initial stage of development of the RTI for $b=1.4$. Triggered by the initial infinitesimal perturbation of the interface with wavelength $\approx13a_z$ corresponding to the maximal wavelength for given $\gamma$, $R_0$ and $b$, the BECs start counter-flowing driven by the magnetic force $b$. The relative shear motion is generated on the sides of the growing perturbations (one of them is shown by arrow), which leads to the KHI and mushroom-shaped bubbles. Coordinates are scaled by $a_z$, time by $(\pi\nu_z)^{-1}$, and density by $n_0$.}
\end{figure}
We use x axis along the interface, z axis perpendicular to the interface plane, Fig. 1 (a).
The trap frequencies $\nu_z=\nu_x=\nu_y/100=100 \mathrm{Hz}$. We also use the "stripe" geometry with $\nu_x=0$, $\nu_y=100\nu_z$ in modelling the 2D KHI, and calculations of the 2D bubble velocity. In all simulations we use 2D equations of motion, where $y$ coordinate does not appear, but the frequency $\nu_y$ is necessary for calculation of the 2D interaction parameters $g_{ij}$ \cite{PethickSmith}. The boundary conditions along all directions are periodic. In fact, any non-zero driving field removes the periodicity, but since we ensure that number density of BEC is negligible at the edge of computational field compared to the bulk value, the periodic boundary conditions is a reasonable numerical approximation.

The macroscopic wave functions $\psi_j$ ($j=1,2$) are scaled by $\sqrt{n_0}$ where $n_0$ is the atom number density in the trap centre, the unit of time $t$ is $1/\pi\nu_z$, and the unit of length is $a_z\equiv\sqrt{\hbar/2\pi\nu_zm}$, where $m$ is the atomic mass.
The gradient of driving magnetic field magnitude $B{'}\equiv dB_x/dz\sim0.6 [\mathrm{Gauss}/\mathrm{cm}]$ is represented by the dimensionless parameter $b(t)=B{'}(t)\mu_Ba_z/ 2\pi\nu_z\hbar$. The dimensionless repulsion parameter is $\gamma\equiv(g_{12}-g)/g\approx0.01$, where $g=g_{11}=g_{22}$ with ${{g}_{jk}}=2\sqrt{2\pi }{{\hbar }^{2}}{{a}_{jk}}/(m{{a}_{y}})$ in 2D, where $a_y\equiv\sqrt{\hbar/2\pi\nu_zm}$ is the oscillator length of the trap in the strongly confined direction \cite{PethickSmith}. We introduce the measure of non-linearity of the system $R_0=\sqrt{gn_0/\hbar\pi\nu_z}$. The parameter $R_0$ determines the Thomas-Fermi radius of the system as $R_{TF}=a_zR_0$, the healing length of the condensate as $\xi=a_z/R_0$, and the sound speed in the trap centre $c_s=\sqrt{2}\pi a_z\nu_zR_0$. The characteristic width of the interface between two BECs is $\Delta_{int}=\xi/\sqrt{\gamma}$ \cite{PRA_86_023614_2012}. The dimensionless 2D GP equations are:
\begin{eqnarray}
\label{GPE}
i{{\partial }_{t}}{{\psi }_{j}}=-{{\nabla }^{2}}{{\psi }_{j}}+V_j{{\psi }_{j}}+R_{0}^{2}\left[ {{\left| {{\psi }_{j}} \right|}^{2}}+\left( 1+\gamma  \right){{\left| {{\psi }_{3-j}} \right|}^{2}} \right]{{\psi }_{j}},
\end{eqnarray}
where $j=1,2$, and $V_j=-1+{{z}^{2}}+\alpha{{x}^{2}}+{{\left( -1 \right)}^{j}}bz$. The parameter $\alpha\equiv(\nu_x/\nu_z)^2$ determines the trap geometry: we use $\alpha=1$ in symmetric trap and $\alpha=0$ in the stripe geometry. The first term in definition of $V_j$ comes from the chemical potentials, and ensures time-independent stationary states. Dimensional form of the term $V_j$ reads $V_j=-\mu_j+m(2\pi\nu_z)^2z^2/2+m(2\pi\nu_x)^2x^2/2+(-1)^{j}z\mu_B B'/2$. In Eq. (1), positive $b$ exerts force along $z$ axis on BEC 1, and the opposite force on BEC 2. The dimensionless parameters $\gamma$, $R_0$ and $b$ comprise the full set of parameters of a binary BEC with symmetric components (equal atom masses, intra-component scattering lengths, and particle numbers).

\subsection{Generation of turbulence}
Dispersion relation for waves on the interface between two weakly repulsive BECs is given by [26]
\begin{equation}
\label{RTI}
\omega_{RT}^2=\sigma k_x^3 - g(t) k_x,
\end{equation}
where $\sigma=\sqrt{\gamma}R_0(\pi\nu_z)^2a_z^3$, and $g(t)=2b(t)(\pi\nu_z)^2a_z$. Note that the form of $\sigma$ below Eq. (\ref{RTI}) is applicable to superfluids with weak intercomponent repulsion. The form of Eq. (\ref{RTI}) is very similar to dispersion relation of capillary-gravitational waves in classical fluids:
\begin{equation}\label{RTIclass}
\omega_{cl}^2=\frac{\sigma_{cl}}{\rho_1+\rho_2} k_x^3 - \frac{\rho_2-\rho_1}{\rho_1+\rho_2}g_{cl} k_x,
\end{equation}
where $\omega_{cl}$ and $k_x$ are frequency and wavenumber, $\sigma_{cl}$ is the interface tension between fluids, $\rho_1$ and $\rho_2$ are mass densities of the lower and the upper fluids, $g_{cl}$ is the gravitational acceleration. Therefore, interface tension in weakly repulsive binary BEC is $\propto\sqrt{\gamma}$. In Fig. 2 we show the dispersion relation, Eq. (\ref{RTI}) for two typical set of parameters $R_0$ and $b$, which displays the unstable long-wavelength region of spectrum characteristic to the RTI. As seen from Eqs. (\ref{RTI}),(\ref{RTIclass}), the role of gravity is played by the magnetic driving force $b$.
\begin{figure}
\includegraphics[width=3.5in]{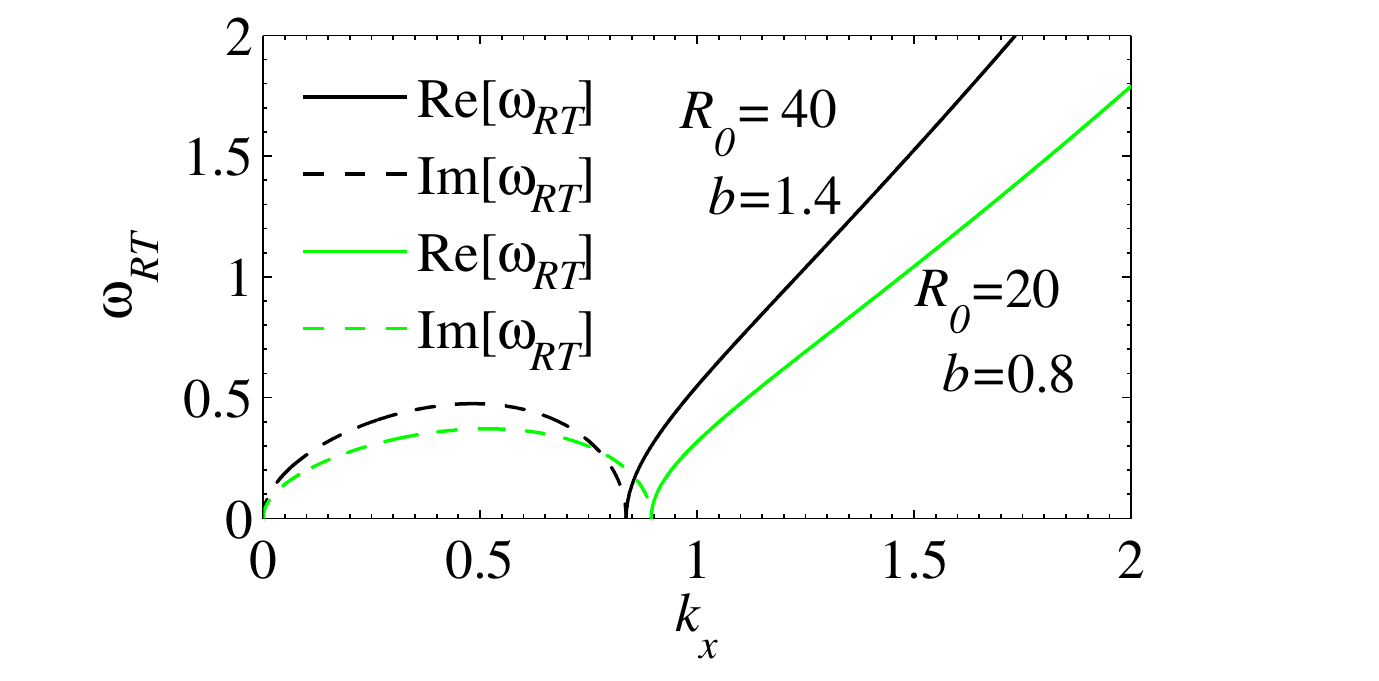}
\caption{(Color online) Frequency $\omega_{RT}$ for waves on the interface between two weakly repulsive BECs for $\gamma=0.01$, and two sets of parameters $R_0$ and $b$, as function of wavenumber $k_x$ of the perturbation. Dashed lines show imaginary $\omega_{RT}$ corresponding to the RTI, solid lines show $\omega_{RT}$ of the quantum "capillary-gravitational" oscillations. Wavenumbers are scaled by $a_z^{-1}$, and frequencies by $2\pi\nu_z$.}
\end{figure}

The two condensates are prepared in the ground state, and the driving force is abruptly turned on at $t=0$. The RTI is initiated at $t=0$ by setting a single-mode infinitesimal perturbation with wavenumber $k_x$ equal to wavenumber of the fastest mode $k_{max}=\sqrt{3}k_{c}$, where $k_c$ and $k_{max}$ are found from ${\omega_{RT}(k_x)}|_{k_x=k_c}=0$, and ${d\omega_{RT}/dk_x}|_{k_x=k_{max}}=0$. The interface perturbation starts to grow due to the RTI, producing layers of the binary superfluid with shear motion, shown by the arrow in Fig. 1 (b). In contrast to single-component superfluids where fluid layers cannot have shear motion, here it is possible due to two-component character of the superfluid. The relative shear motion leads to the KHI which rolls the interface producing vortex bundles and mushroom-shaped bubbles, see Fig. 3 (a) and Sec. IV.
\begin{figure*}
\includegraphics[width=7.3in]{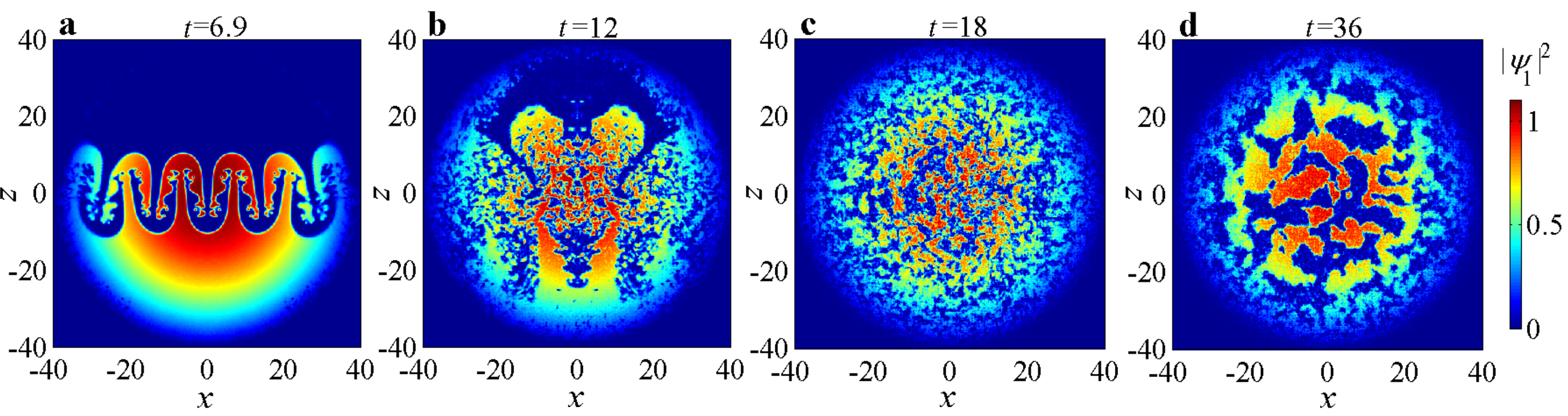}
\caption{(Color online) Binary quantum turbulence generated for $\gamma=0.01$, $R_0=40$ and $b=1.4$. Density of BEC 1 for subsequent instants of time showing generation (a), (b), and free decay of turbulence (c), (d), which follows the initial stage shown in Fig. 1. Coordinates are scaled by $a_z$, time by $(\pi\nu_z)^{-1}$, and density by $n_0$.}
\end{figure*}
The "mushrooms" grow with a characteristic velocity, estimated from the potential flow model \cite{Layzer1955}. The model  predicts in the limit $\rho_1\lesssim\rho_2$ and $\sigma_{cl}=0$, that the asymptotic bubble velocity on late stages of the RTI is given by $U_{b}^{(2Dclass)}=\sqrt{g_{cl}/3k_x}$. Comparing Eqs. (\ref{RTI}) and (\ref{RTIclass}) we obtain for 2D bubble in binary BEC
\begin{equation}
\label{Ub}
U_b^{2D}= a_z\pi\nu_z\times[(2b/3\pi)\times(\lambda_x/a_z)]^{1/2},
\end{equation}
where $\lambda_x=2\pi/k_x$ is the dimensional wavelength of the interfacial perturbation.
This estimate agrees well with the numerical simulation in 2D performed in the stripe geometry, as shows Fig. 4.
\begin{figure}
\includegraphics[width=3.45in]{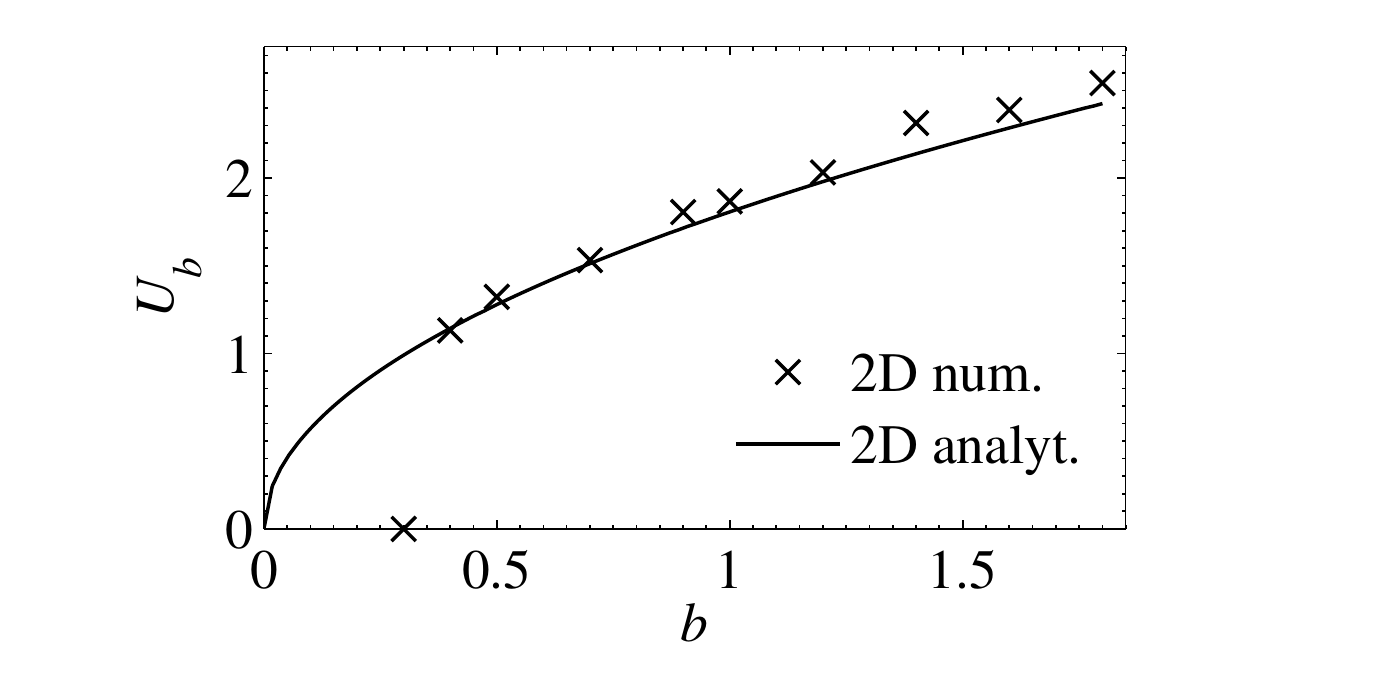}
\caption{Velocity of the bubble tip in units of $\pi a_z\nu_z$ as function of the scaled strength of the driving force for $\gamma=0.01$ and $R_0=30$. Line shows the analytical prediction, Eq. (\ref{Ub}), and markers show the results of numerical simulation.}
\end{figure}

On the late non-linear stage of dynamics, vorticity sheets on the interface are rolled-up by the KHI, spliting each BEC into droplets, and producing non-singular quantum vortex dipoles, which eventually leads to QT, Fig. 3 (b). In the present system vortices are non-singular because their cores are filled by a non-rotating superfluid. The total fluid velocity of a non-singular quantum vortex remains finite unlike in single-component superfluids.

Certain experimental conditions are required to generate QT by the RTI. In our system vortex pairs form from the phase jumps of the condensates at the interfaces with shear flow. However, phase singularities penetrate superfluids with the help of the KHI which corrugates interfaces with shear motion, Fig. 3 (a). As the characteristic wavenumber of the superfluid flow is $q\sim\sqrt{bR_0}$ \cite{PRA_85_013630_2012}, the total phase change along the trap is $R_0\sqrt{bR_0}$. This estimate immediately gives the  number of phase jumps as $R_0\sqrt{bR_0}/2\pi$, in a good agreement with numerical simulations, and may be observed experimentally. Condition for turbulence generation is that there are many vortices in the bulk \cite{LaserPhysLett8_393}, and that the critical wavelength of the RTI is not greater than the system size. It follows from Eq. (\ref{RTI}) that the critical wavelength $\lambda_c$ for triggering the RTI is given by
\begin{equation}
\label{lambdaC}
{\lambda_c}=a_z\sqrt{2}\pi\sqrt{{\gamma^{1/2}}{{{R}}_{0}}/{b}},
\end{equation}
the minimal driving force $b_{min}$ required for the RTI is obtained by setting $\lambda_c=2R_{TF}$, which leads to estimate $b_{min}\approx\pi^2\sqrt{\gamma^{1/2}/R_0}$. Plugging $b_{min}$ into the condition $R_0\sqrt{bR_0}/2\pi\gg1$, we obtain the condition for turbulence generation as
\begin{equation}\label{TURBcondition}
R_0\gamma^{1/4}\gg1.
\end{equation}
Note that the condition of thin interface $R_0/\Delta_{int}\gg1$ also leads to Eq. (\ref{TURBcondition}). The estimate of number of phase jumps determines dependence of the maximal value of the integrated vorticity $\Omega$ as
\begin{equation}\label{maxOmega}
max\{\Omega\}\propto\sqrt{b},
\end{equation}
where $\Omega\equiv\int dz\,dx\,|\pmb{\omega}|$ is the generalization of the vortex line length, and $\pmb{\omega}=\nabla\times[\mathbf{J}/n]$. Here
\begin{equation}
\label{J}
\mathbf{J}=\frac{\hbar }{2mi}\underset{j=1,2}{\mathop \sum }\,\left( \psi _{j}^{*}\pmb{\nabla} {{\psi }_{j}}-{{\psi }_{j}}\pmb{\nabla} \psi _{j}^{*} \right)
\end{equation}
is the total current, and $n=|\psi_1|^2+|\psi_2|^2$ is the total density. The result in Eq. (\ref{maxOmega}) is in a good agreement with our numerical result shown in Fig. 5.
\begin{figure}
\includegraphics[width=3.45in]{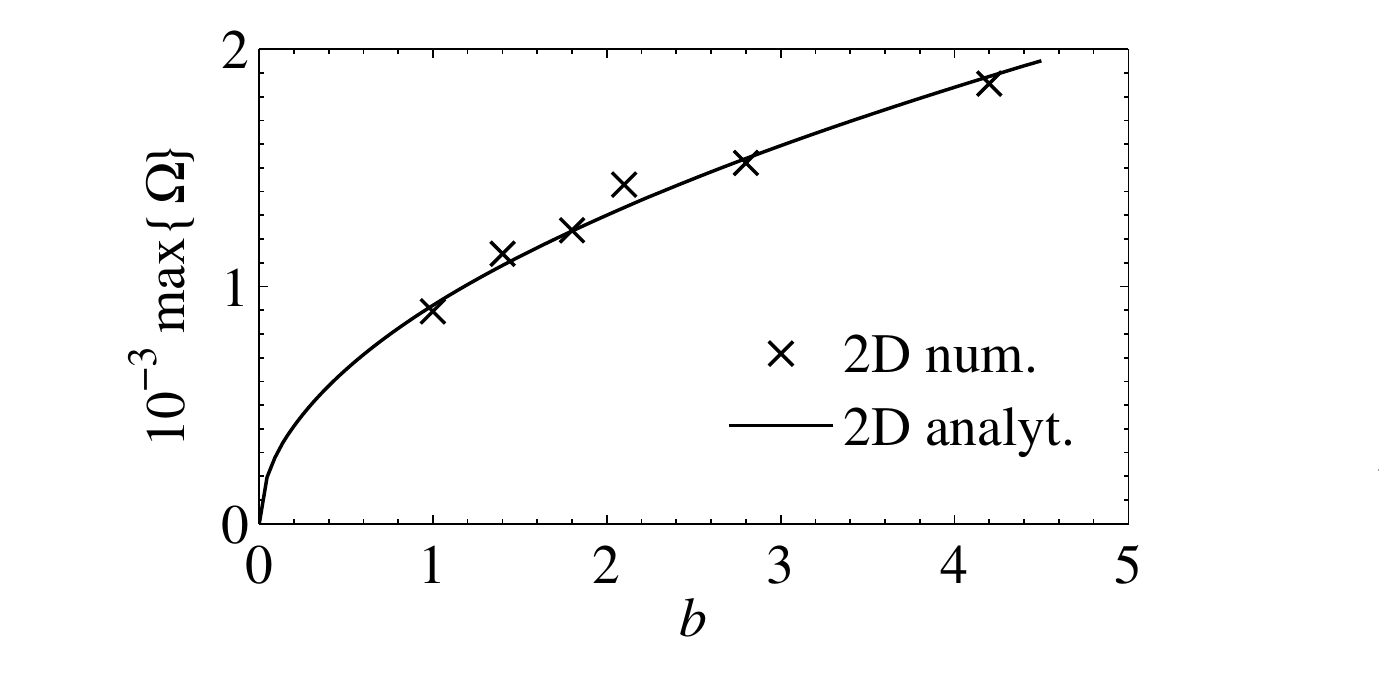}
\caption{Maximum integrated vorticity $\Omega$ in units of $2\pi a_z^2\nu_z$ developed in 2D RTI as function of the scaled driving force $b$, for $\gamma=0.01$ and $R_0=30$. Line shows the analytical prediction, Eq. (\ref{maxOmega}), and markers show the results of numerical simulation.}
\end{figure}

Time $\tau_{turb}$ needed to generate turbulence can be obtained from Eq. (\ref{Ub})
\begin{equation}\label{tauturb}
\tau_{turb}=a_zR_0/U_b.
\end{equation}
To study the freely decaying QT we turn off the driving force at $t=\tau_{turb}$, and show the free QT in Fig. 1 (c), (d). For this numerical simulation we used a 2D grid with 512$\times$512 points. Although the simulation conserves the total energy within less than one percent (and the error oscillates in time giving a constant mean value), the resolution tests show that the error systematically decreases with increase of the grid points number not changing the result of simulations. We then analyse the resulting turbulence.

\section{Kolmogorov scaling of the incompressible kinetic energy and the spin energy}
As observed in numerical simulations, the factor $\gamma$ in definition of the the spin-dependent energy of contact atom-atom
interaction $e_{spin}=-\int dz\,dx\,(g_{12}-g)(n_1-n_2)^2/2=-\gamma\int dz\,dx\, g(n_1-n_2)^2/2$, makes the modulus of the turbulent spin-dependent energy $E_{spin}\equiv|e_{spin}|$ as small as the turbulent kinetic energy $E_k=\int dz\,dx\,m\mathbf{J}^2/n$, because $0<\gamma\ll1$. Thus, the incompressible kinetic energy remains the essential characteristic of turbulence despite of its smallness. As noted in Sec. I, the total fluid velocity of binary weakly repulsive BEC is expected to be almost incompressible, which is in agreement with the numerical results. To calculate the energy spectrum of the resulting states (Figs. 3 (c), (d)) we consider the incompressible part of the quantum kinetic energy inside the Thomas-Fermi raduis. First the quantum pressure is dropped from the kinetic energy $\int dz\,dx\,|\pmb{\nabla}\psi_1|^2+|\pmb{\nabla}\psi_2|^2$, where we also omit the dimensional factors for simplicity, yielding  $\int dz\,dx\,[\mathbf{w}_1(\mathbf{r})]^2+[\mathbf{w}_2(\mathbf{r})]^2$, where $\mathbf{w}_j(\mathbf{r})\equiv\sqrt{n_j}\pmb{\nabla}\,\phi_j$, with $n_j=|\psi_j|^2$, and $\phi_j=(2i)^{-1}\ln(\psi_j/psi_j^*)$.  The angle-averaged incompressible kinetic energy $E_{kin}^{(\mathrm{i})}$ is then given by
\begin{equation}\label{Ek}
E_{kin}^{(\mathrm{i})}\left( k \right)=\underset{k<|\mathbf{k} |<k+dk}{\mathop \sum }\,{{e}^{(\mathrm{i})}}(\mathbf{k}),
\end{equation}
where $e^{(\mathrm{i})}(\mathbf{k})=|\mathbf{w}_1^{(\mathrm{i})}(\mathbf{k})|^2+|\mathbf{w}_2^{(\mathrm{i})}(\mathbf{k})|^2$. The vector fields $\mathbf{w}_j^{\mathrm{i}}(\mathbf{k})$ are the incompressible parts of the Fourier images $\mathbf{w}_j(\mathbf{k})$ corresponding to the fields $\mathbf{w}_j(\mathbf{r})$. Generally, the incompressible part is given by $\mathbf{w}_j^{\mathrm{i}}(\mathbf{k})=\mathbf{w}_j(\mathbf{k})-\mathbf{k}[\mathbf{k}\cdot \mathbf{w}_j(\mathbf{k})]/|\mathbf{k}|^{2}$, because its inverse Fourier image $\mathbf{w}_j^{\mathrm{i}}(\mathbf{r})$ satisfies $\pmb{\nabla}\cdot\mathbf{w}_j^{\mathrm{i}}(\mathbf{r})\equiv0$, and $\pmb{\nabla}\times[\mathbf{w}_j(\mathbf{r})-\mathbf{w}_j^{\mathrm{i}}(\mathbf{r})]\equiv0$.

Figure 6 shows that the Kolmogorov spectrum indeed is present in turbulence of immiscible superfluids with weak interspecies repulsion $\gamma\gtrsim0$.
\begin{figure}
\includegraphics[width=3.45in]{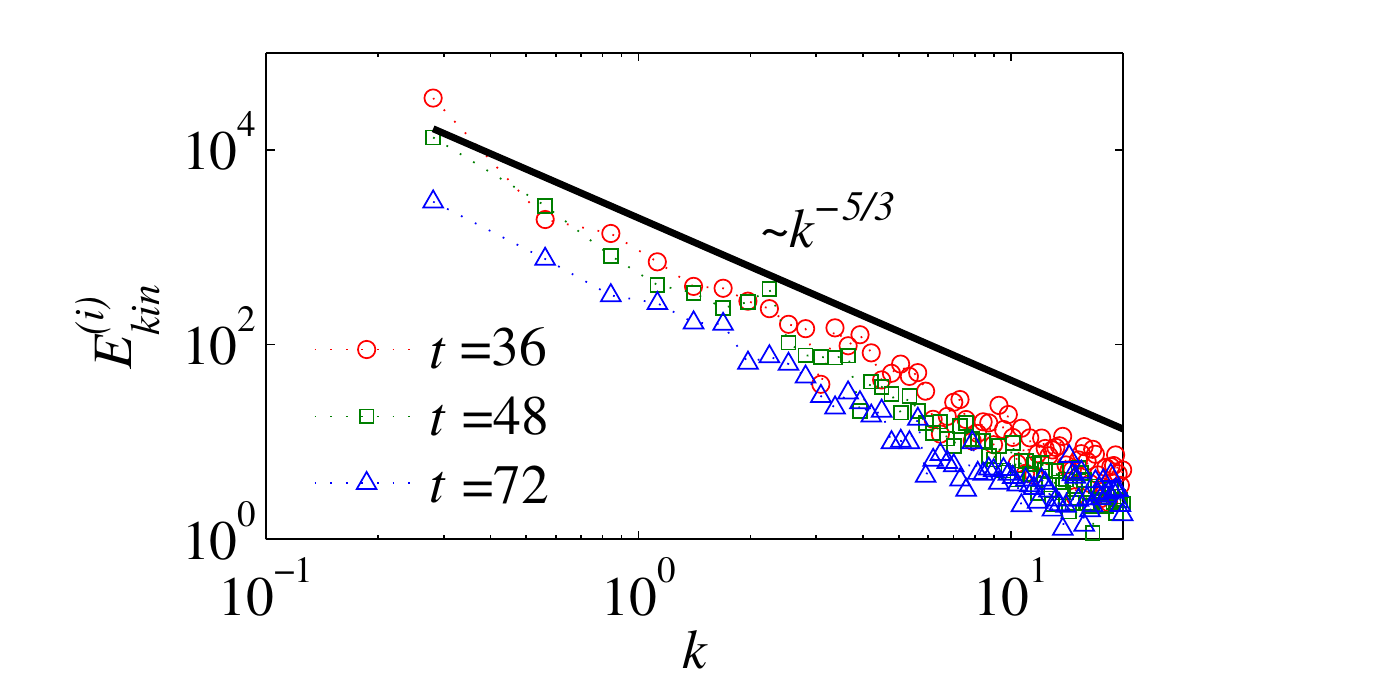}
\caption{(Color online) Angle-averaged spectrum of incompressible kinetic energy of binary quantum turbulence for $\gamma=0.01$, $R_0=40$ and $b=1.4$, corresponding to Fig. 3 (c), (d) and an intermediate instant. Straight lines show analytical predictions. Wavenumbers are scaled by $a_z^{-1}$, and energy density by $\pi\hbar\nu_zn_0$.}
\end{figure}
We plot the spectrum only in the inertial range corresponding to length scales from the dynamical viscous length scale up to the intergral length scale. The dynamical viscous length scale is given by the vortex core size $\sim\Delta_{int}=a_z/R_0\sqrt{\gamma}$. Importantly, for weakly repulsive BECs the viscous scale is larger than the healing length by a factor of $1/\sqrt{\gamma}$, which distinguishes weakly repulsive BECs from single-component superfluids and strongly repulsive or attractive BECs. As the integral scale we take $a_zR_0/2$ because it corresponds to almost homogenous total density region in the trap centre. It is interesting to note that in contrast to one-component GP turbulence \cite{PRL_94_065302_2005}, no thermal dissipation is needed in our system to obtain the Kolmogorov spectrum because, first, the dynamical viscous scale is larger than the healing length where the thermal dissipation is important \cite{PRL_97_145301_2006}, and second, because the kinetic energy is just a small fraction of the total energy. The latter fact implies that the "large" energies associated with trap and non-linear interactions act as a "reservoir" shifting the dissipative length scale.

In addition to the universal Kolmogorov scaling of the incompressible kinetic energy, turbulence in superfluids with spin degrees of freedom exhibits a self-similar cascade of the spin-dependent energy of contact atom-atom
interaction. Two-component GP equations may be represented as a pseudospin system \cite{PRA_86_023614_2012}, and therefore the universal spin-energy scaling $E_{spin}\sim k^{-7/3}$ recently proposed for spinor turbulent Bose-Einstein condensates in the ferromagnetic phase \cite{PRA_85_033642_2012} is expected to hold as well. We check this by angle-averaging the dimensionless spin-dependent energy density $E_{spin}=\gamma R_0^2(|\psi_1|^2-|\psi_2|^2)^2/2$ in Fourier space similarly to Eq. (\ref{Ek}), and plot the result in Fig. 7.
\begin{figure}
\includegraphics[width=3.45in]{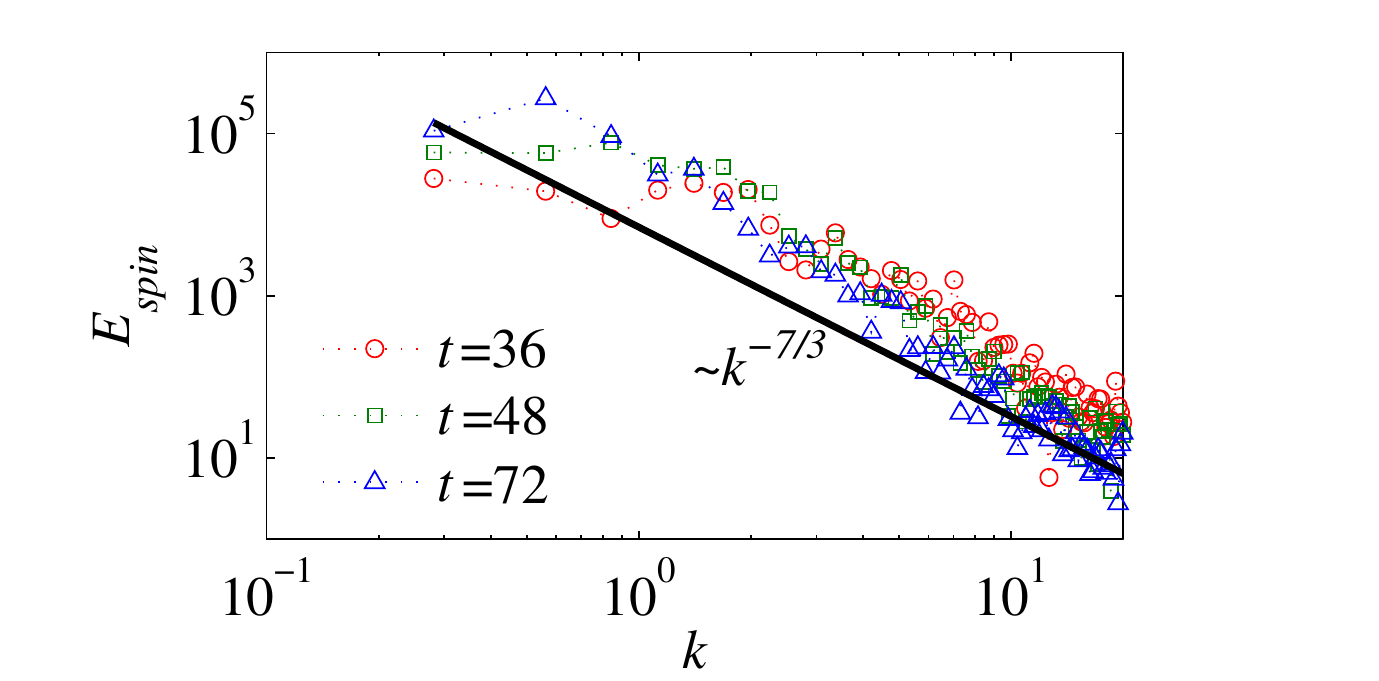}
\caption{(Color online) Angle-averaged spectra of spin-dependent interaction energy of binary quantum turbulence for $\gamma=0.01$, $R_0=40$ and $b=1.4$, corresponding to Fig. 3 (c), (d) and an intermediate instant. Straight lines show analytical predictions. Wavenumbers are scaled by $a_z^{-1}$, and energy density by $\pi\hbar\nu_zn_0$.}
\end{figure}
We find that the numerical results are in reasonable agreement with the analytical prediction; a fit to the numerical data yields a power-law $k^{-3}$. Presumably the slightly larger slope in our case is due to the parameter $\gamma$ being much smaller than in \cite{PRA_85_033642_2012}.

\section{Generation of vortex bundles by the KHI}
To reveal the physics behind the Kolmogorov spectrum formation in our system, we further consider the basic process of the quantum vorticity generation on the sides of the RT bubbles, the KHI. One of these sides is marked by an arrow in Fig. 1 (b), where the relative shear motion is along a locally flat interface. To get insight into the mechanism of rolling-up of the interface on the later stage of the RTI, we then consider a model problem - the KHI in a two-dimensional harmonic trap unconfined in the direction of motion (along $x$ axis, \emph{i.e.} $\alpha=0$) and non-stratified ($B'=0$).

In classical fluid dynamics the KHI is instability of the interface between the fluid layers with relative shear motion. In superfluids such situation is possible only for a mixture, because velocity field of each superfluid is potential. To initiate the shear flow along the interface, we multiply the steady ground-state wavefunction of component $j$ by the factor $\exp[-i(-1)^jq_0x]$, where $q_0=2.5a_z^{-1}$, and simulate dynamics for two systems which differ only by size along the interface, Fig. 8.

Surface tension between fluid components tends to increase the critical velocity for triggering the KHI, and may completely suppress the roll-up of the vorticity sheet. This has been studied for $1+\gamma=10$ in \cite{PRB_81_094517_2010}. On the contrary, our system is a weakly repulsive binary BEC, $\gamma=0.01$. This leads to the fundamental difference of the present QT and the QT studied previously: here the quasi-classical roll-up of the vorticity sheet is favorable and represents a quantum counterpart of the basic process of generation of CT.

Similarly to the 2D CT where vorticity accumulates on the mode with the largest possible wavelength, the KHI in our 2D system with weakly repulsive intercomponent interaction $0<\gamma\lesssim1$, produces a bundle of quantum vortices, with the largest possible size, Figs. 8 (b), (d). Note that it takes longer to generate the bundle in the wider configuration (Figs. 8 (c), (d)).
\begin{figure}
\includegraphics[width=3.65in]{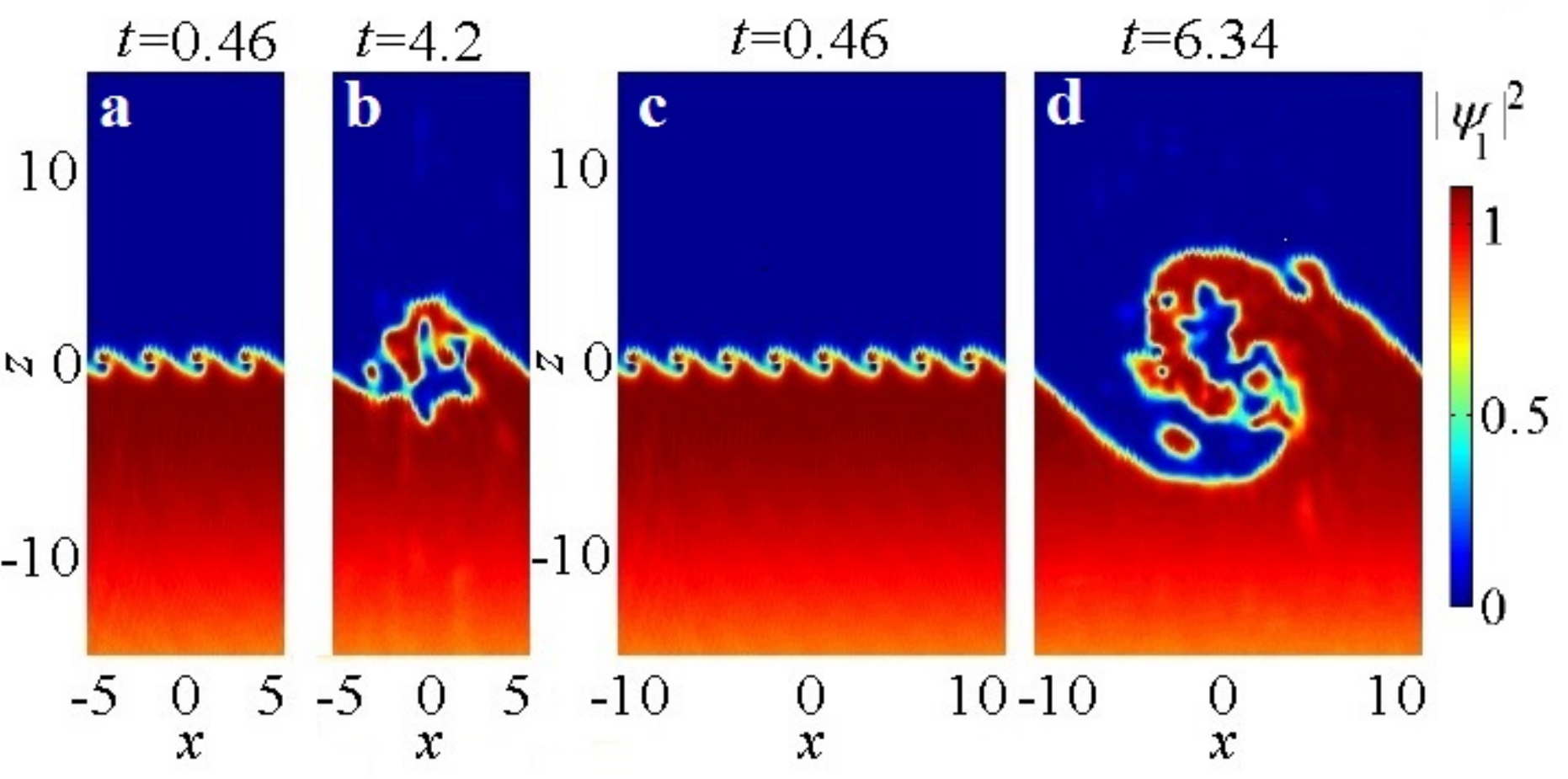}
\caption{(Color online) Development of the KHI in two-component BEC with a shear flow for $\gamma=0.01$ and $R_0=30$. The shear flow is between BEC at $z<0$ (along $x$), and BEC at $z>0$ (opposite to $x$). Panels (a), (b) show BEC 1 in a system which is twice narrower in $x$ direction than the system shown in (c), (d), while the other parameters are the same. BEC 2 is located symmetrically and not shown. The unstable mode is the same for both systems,  (a), (c). At later times (b), (d), the sheet with vorticity (the interface) rolls up transferring vorticity to the largest possible eddy and forming a vortex bundle. Coordinates are scaled by $a_z$, time by $1/\pi\nu_z$, and density by $n_0$..}
\end{figure}
More importantly, generation of bundles in weakly-repulsive superfluids makes the late stage of the KHI very different from the KHI in superfluids with stronger repulsion considered e.g. in \cite{PRB_81_094517_2010}. This is the reason why the Kolmogorov spectrum is so pronounced in our system.

A very convenient feature of our system is possibility to generate binary quantum turbulence with arbitrarily low Mach numbers. The Mach number characteristic to our system is
\begin{equation}
\mathrm{Ma}\sim2\sqrt{bR_0}/\tilde{c}_s=\sqrt{2b/R_0},
\end{equation}
where $2\sqrt{bR_0}$ is the characteristic dimensionless shear flow velocity induced by the magnetic driving force (discussed in Sec. II a), and $\tilde{c}_s=\sqrt{2}R_0$ is the dimensionless sound speed. Using Eq. (\ref{lambdaC}) we find for typical $R_0\sim30$ that $b\sim0.5$, and thus $\mathrm{Ma}\lesssim0.2$ in Figs. 1, 3, 8. Using the value $b_{min}$ for given $R_0$, we find that the Mach number scales as $\mathrm{Ma}\propto\gamma^{1/4}/R_0$, and may be made arbitrarily small by taking larger $R_0$.

Owing to low Mach number, the total fluid dynamics is to a good approximation incompressible, while dynamics of each of the condensates is compressible because within the TF radius of the total fluid each component has large regions of low density. Vortices appear and annihilate by pairs leading to non-conservation of the enstrophy in the 2D system. We find that soon after the instant of switch-off of the driving field, the integrated vorticity $\Omega$ decreases in time which shows that our two-dimensional system exhibits unique properties \cite{PhysRep_362_1_2002}: being incompressible (its total density is almost unaffected by dynamics) it demonstrates essential features of compressible 2D dynamics (annihilation of vortex pairs - enstrophy non-conservation).

\section{Conclusion}
In conclusion, we have studied for the first time QT generated by the RTI and KHI, and showed that the Kolmogorov scaling law holds in weakly repulsive immiscible QT. The structure of topological charge (vorticity) in weakly repulsive binary superfluids is different from previously studied types in superfluid turbulence. In 2D CT vorticity resides on 2D structures, which is impossible for vorticity in 2D single-component superfluids. Dynamics of shear layers in weakly repulsive binary superfluids is remarkably similar to dynamics of shear layers in CT, in contrast to immiscible superfluids with strong intercomponent repulsion, and miscible superfluids. Our method of stirring of quantum fluids is advantageous for the cases when thermal and compressible excitations are undesirable. The Mach number can be made arbitrarily small, which allows avoiding thermal excitations in the system, in contrast to previously studied QT. Our system effectively generates QT from a single-mode RTI, which remains a challenging problem even in classical RTI \cite{Phys_Fluids_16_1668_2004}, \cite{Phys_Fluids_24_074107_2012}, thus providing a new intersection between QT and CT. The present system may be conveniently realized experimentally which opens a novel avenue for studies of turbulence.

{\bf Acknowledgments} D.K. is grateful to Prof. C. J. Pethick for useful discussions. This work was supported by the Swedish Research Council (VR), by the Kempe foundation, and by the Baltic Donation foundation.

\end{document}